\documentclass[conference]{IEEEtran}
\def\BibTeX{{\rm B\kern-.05em{\sc i\kern-.025em b}\kern-.08em
    T\kern-.1667em\lower.7ex\hbox{E}\kern-.125emX}}
\usepackage{amsmath}
\usepackage{amsthm}
\usepackage{amsfonts}
\usepackage{amssymb}
\usepackage{graphicx}
\usepackage{cite}
\usepackage{subfigure}
\usepackage{epstopdf}
\usepackage{balance}
\usepackage{algorithm}
\usepackage{algorithmic}
\usepackage{enumerate}
\usepackage{color}
\usepackage{array}
\usepackage{indentfirst}
\usepackage{multirow}
\usepackage{booktabs}
\usepackage{color}
\usepackage{slashbox}
\usepackage{diagbox}
\usepackage{threeparttable}
\usepackage[svgnames,table]{xcolor}
\usepackage{bm}
\usepackage{subfigure}

\makeatletter

\newcommand{\Rmnum}[1]{\expandafter\@slowromancap\romannumeral #1@}
\makeatother

\begin{document}
\title{ Deep Transfer Learning-Assisted Signal Detection for Ambient Backscatter Communications\vspace{-0.6cm}}

\author{Chang Liu$^{\ast}$, Xuemeng Liu$^{\S}$, Zhiqiang Wei$^{\ast}$, Derrick Wing Kwan Ng$^{\ast}$, \\ Jinhong Yuan$^{\ast}$, and $^{\dag}$Ying-Chang Liang \\
$^{\ast}$School of Electrical Engineering and Telecommunications, the University of New South Wales, Australia\\
$^{\S}$Dalian University of Technology, China\\
$^{\dag}$University of Electronic Science and Technology of China, China \\
Email: $^{\ast}$\{chang.liu19, zhiqiang.wei, w.k.ng, j.yuan\}@unsw.edu.au, $^{\S}$xuemeng.liu.ac@gmail.com, 
$^{\dag}$liangyc@ieee.org \vspace{-0.2cm}}

\maketitle

\begin{abstract}
Existing tag signal detection algorithms inevitably suffer from a high bit error rate (BER) due to the difficulties in estimating the channel state information (CSI). To eliminate the requirement of channel estimation and to improve the system performance, in this paper, we adopt a deep transfer learning (DTL) approach to implicitly extract the features of communication channel and directly recover tag symbols. Inspired by the powerful capability of convolutional neural networks (CNN) in exploring the features of data in a matrix form, we design a novel covariance matrix aware neural network (CMNet)-based detection scheme to facilitate DTL for tag signal detection, which consists of offline learning, transfer learning, and online detection. Specifically, a CMNet-based likelihood ratio test (CMNet-LRT) is derived based on the minimum error probability (MEP) criterion.
Taking advantage of the outstanding performance of DTL in transferring knowledge with only a few training data, the proposed scheme can adaptively fine-tune the detector for different channel environments to further improve the detection performance. Finally, extensive simulation results demonstrate that the BER performance of the proposed method is comparable to that of the optimal detection method with perfect CSI.
\end{abstract}

\section{Introduction\label{sect: intr}}
With the rapid development of the fifth-generation (5G) wireless communications, the Internet-of-Things (IoT) has been proposed as a versatile system which enables the connections of massive devices through the internet \cite{palattella2016internet, wong2017key}. However, one of the key challenges in realizing the era of IoT is the restricted lifetime of networks, as only limited energy storage capacity in batteries can be adopted for most IoT devices \cite{wu2017overview}. To overcome this problem, one of the promising solutions is ambient backscatter communication (AmBC), which enables passive backscatter devices (e.g., tags, sensors) to transmit their information bits to dedicated readers over ambient radio-frequency (RF) signals (e.g., Wi-Fi signals, cellular base station signals, and TV tower radio signals) \cite{van2018ambient}. In an AmBC system, an AmBC tag could transmit its binary tag symbols by choosing whether to backscatter the ambient RF signals or not. Thus, one of the key tasks of an AmBC system is to perform tag signal detection, i.e., recovering the tag signal at the reader, which has attracted tremendous attention from both academia \cite{van2018ambient} and industry \cite{hoang2020ambient}, respectively.
Generally, there are two main challenges for tag detection: (1) since both the direct link signal from the RF source and the backscatter link signal from the tag could be received by the reader simultaneously, the received direct link signal generally causes severe interference to the received backscatter link signal; (2) in contrast to the traditional wireless communication systems, estimating the channel state information (CSI) in AmBC systems is challenging due to the lack of pilot signals sent from the ambient RF source \cite{liu2020deepresidual}.

Recently, various effective algorithms have been proposed for tag signal detection in AmBC systems. In \cite{lu2015signal}, an energy detection (ED) method was proposed to decode tag symbols which achieves good detection performance. However, this energy detection method requires the knowledge of perfect CSI which is not available in practical AmBC systems. To overcome this issue, a semi-coherent detection method was designed in \cite{qian2017semi} which requires only a few pilots and unknown data symbols. In order to eliminate the process of channel estimation, Wang \emph{et al.} \cite{wang2016ambient} adopted a differential encoding scheme for tag bits and proposed a minimum BER detector. On top of \cite{wang2016ambient}, a fundamental study of the BER performance for non-coherent detectors was conducted in \cite{qian2016noncoherent}.
Furthermore, machine learning (ML)-based methods have been proposed recently which aim to directly recover the tag signals without the need of estimating relevant channel parameters explicitly. For example, Hu \emph{et al.} \cite{hu2019machine} transformed the task of tag signal detection into a classification task and designed a support vector machine (SVM)-based energy detection method to improve the BER performance. However, the proposed ML-based method requires a large number of training pilots which reduces the system effective throughput and is not suitable for communication systems with short coherence times. More importantly, there exists a large gap between the proposed method and the optimal method. As a result, considering the time-varying nature of wireless systems, a more practical detector, which can dynamically adapt itself to the changes of channel environment, is expected.

In contrast to the traditional ML methods \cite{liu2019deepcnn, liu2020deepresidualL}, deep transfer learning (DTL), which adopts a deep neural network (DNN) to extract the time-varying features with a few online training data by transferring knowledge from a source domain to a target domain, has proven its powerful capability in capturing time-varying features in numerous research fields, cf. \cite{pan2009survey, tan2018survey, liu2020deep}.
Motivated by this, in this paper, we propose a DTL approach to capture the real-time features of channel environment to further improve the tag signal detection performance for AmBC systems. The main contributions of this paper are as follows:
\begin{enumerate}[(1)]
\item In contrast to the traditional detection methods requiring explicit channel estimation, e.g., \cite{qian2017semi, wang2016ambient, qian2016noncoherent}, we propose a DTL approach for tag signal detection to implicitly extract the features of channel and directly recover tag symbols, which adopts a DNN to transfer the knowledge learnt from one tag detection task under offline channel coefficients to another different but related tag detection task in real-time.
\item We creatively adopt a convolutional neural network (CNN) to explore the features of the sample covariance matrix \cite{liu2019deep} and design a DTL-oriented covariance matrix aware neural network (CMNet) for tag signal detection. Exploiting the powerful capability of CNN in exploring features of data in a matrix form, the proposed method could further improve the BER performance.
\item Specifically, according to the minimum error probability (MEP) criterion, a CMNet-based likelihood ratio test (CMNet-LRT) is derived for tag signal detection, which enables the design of an effective detector.
\end{enumerate}

The remainder of this paper is organized as follows. Section \Rmnum{2} formulates the AmBC system model. In Section \Rmnum{3}, a CMNet-based DTL scheme and the related algorithm are proposed for tag signal detection. Extensive simulation results are presented to verify the efficiency of the proposed method in Section \Rmnum{4}, and Section \Rmnum{5} finally concludes the paper.

\emph{Notations}: Superscripts $T$ and $H$ indicate the transpose and conjugate transpose, respectively. Term ${\mathcal{CN}}( \bm{\mu},\mathbf{\Sigma} )$ represents a circularly symmetric complex Gaussian (CSCG) distribution with mean vector $\bm{\mu}$ and covariance matrix $\mathbf{\Sigma}$. Term ${\bf{I}}_M$ is used to denote the $M$-by-$M$ identity matrix and ${\mathbf{0}}$ is used to denote the zero vector. $(\cdot)^{-1}$ indicates the matrix inverse operation. $\det(\cdot)$ is the determinant operator. $E(\cdot)$ represents the statistical expectation. $\|\cdot\|^2$ denotes the norm of an input vector. $\exp(\cdot)$ represents the exponential function. $\mathbb{C}$ denotes the set of complex numbers.

\begin{figure}[t]
  \centering
  \includegraphics[width=2in, height=1in]{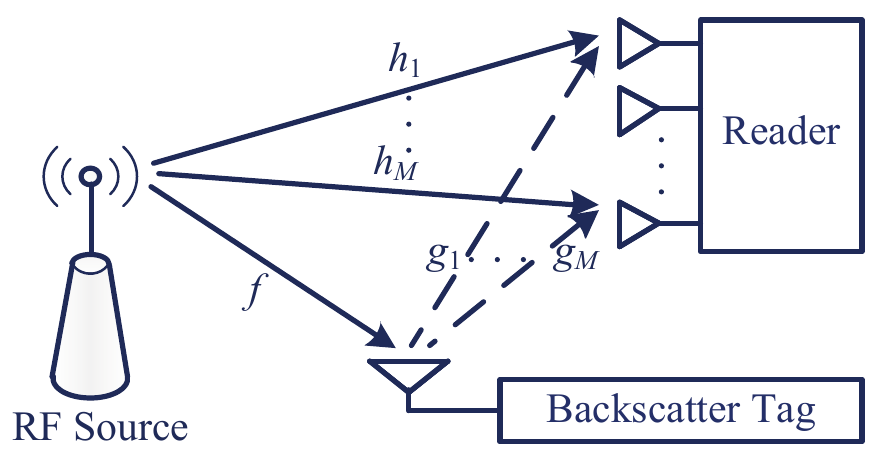}\vspace{-0.2cm}
  \caption{ An illustration of the considered AmBC system. }\vspace{-0.45cm}
\end{figure}

\section{System Model}
As depicted in Fig. 1, in this paper, we consider a general AmBC system, which consists of an ambient RF source, a backscatter tag, and a reader. The reader is equipped with an $M$-element antenna array for signal detection, meanwhile the RF source and the passive tag are both equipped with a single-antenna. Due to the broadcasting nature of the RF source, the transmitted RF signal is received by both the reader and the tag simultaneously. The passive device, tag, can then transmit its binary modulated tag symbols by choosing whether to reflect the ambient RF signals to the reader. In this case, the reader is able to recover the tag symbols through sensing the changes of the received signals.

The frame structure of the received signal at the reader is illustrated in Fig. 2, each frame consists of $P$ pilot symbols and $T-P$ $(T>P)$ data symbols. The pilot tag symbols are known by the reader, and the remaining tag symbols are used for data transmission. In the considered AmBC system, the tag transmits its information bits at a rate $N$ times lower than the sampling rate of the RF source signal. Thus, we define the source-to-tag ratio (STR) as the number of RF source symbols in one tag symbol period, which is $N$ in the considered system.
Denote by $c^{(t)}\in \{0,1\}$ the $t$-th tag symbol with the binary on-off keying modulation, i.e., $c^{(t)} = 0$ refers that the tag does not reflect the RF source signal; otherwise, the tag reflects the RF source signal. Correspondingly, we use $s_n^{(t)}$ to denote the $n$-th RF source signal sample within the tag symbol $c^{(t)}$.
Denote by $ \mathbf{x}_n^{(t)}={{[x_{n,1}^{(t)},x_{n,2}^{(t)},\cdots ,x_{n,M}^{(t)}]}^{T}}, n \in \{ 0,1,\cdots,N-1\}$, the $n$-th observation vector at the reader within the $t$-th, $t\in \{ 1,\cdots,T \}$, tag symbol period, where $x_{n,m}^{(t)}$, $m \in \{1,2, \cdots, M\}$, indicates the $n$-th discrete-time sample observed at the $m$-th antenna element.
In this case, the received signal at the reader can be expressed as
\begin{equation}\label{x_n_t}
\mathbf{x}_n^{(t)} = \mathbf{h}s_n^{(t)} + \alpha f\mathbf{g}s_n^{(t)}c^{(t)} + \mathbf{u}_n^{(t)}, \forall n,t.
\end{equation}
Here, $\mathbf{h}=[h_1, h_2, \cdots, h_M ]^T$ is the direct link channel coefficient vector, where $h_m \in \mathbb{C}$ is the channel coefficient from the RF source to the $m$-th antenna at the reader. Correspondingly, $\mathbf{g}=[g_1, g_2, \cdots, g_M]^T$ is the backscatter link channel coefficient vector, where $g_m \in \mathbb{C}$ is the channel coefficient from the tag to the $m$-th antenna at the reader. Variables $f$, $\alpha\in \mathbb{C}$ represent the channel coefficient from the RF source to the tag and the reflection coefficient of the tag, respectively. Considering that the ambient source signal in practice may arise from an unknown or indeterminate ambient RF source, a general approach is to model $s_n^{(t)}$ by a CSCG random variable \cite{qian2017semi, wang2016ambient, qian2016noncoherent}, i.e., $s_n^{(t)}\sim \mathcal{C}\mathcal{N}(0,\sigma_s^2)$, where $\sigma_s^2$ is the signal variance.
In addition, $\mathbf{u}_n^{(t)}\in \mathbb{C}^{M\times1}$ denotes the noise vector and is assumed to be an independent and identically distribution (i.i.d.) CSCG random vector with $\mathbf{u}_n^{(t)}\sim \mathcal{CN}( \mathbf{0},\sigma _u^2{{\mathbf{I}}_M} )$, where $\sigma_u^2$ is the noise variance at each antenna of the reader.

\begin{figure}[t]
  \centering
  \includegraphics[width=2.6in, height=1.in]{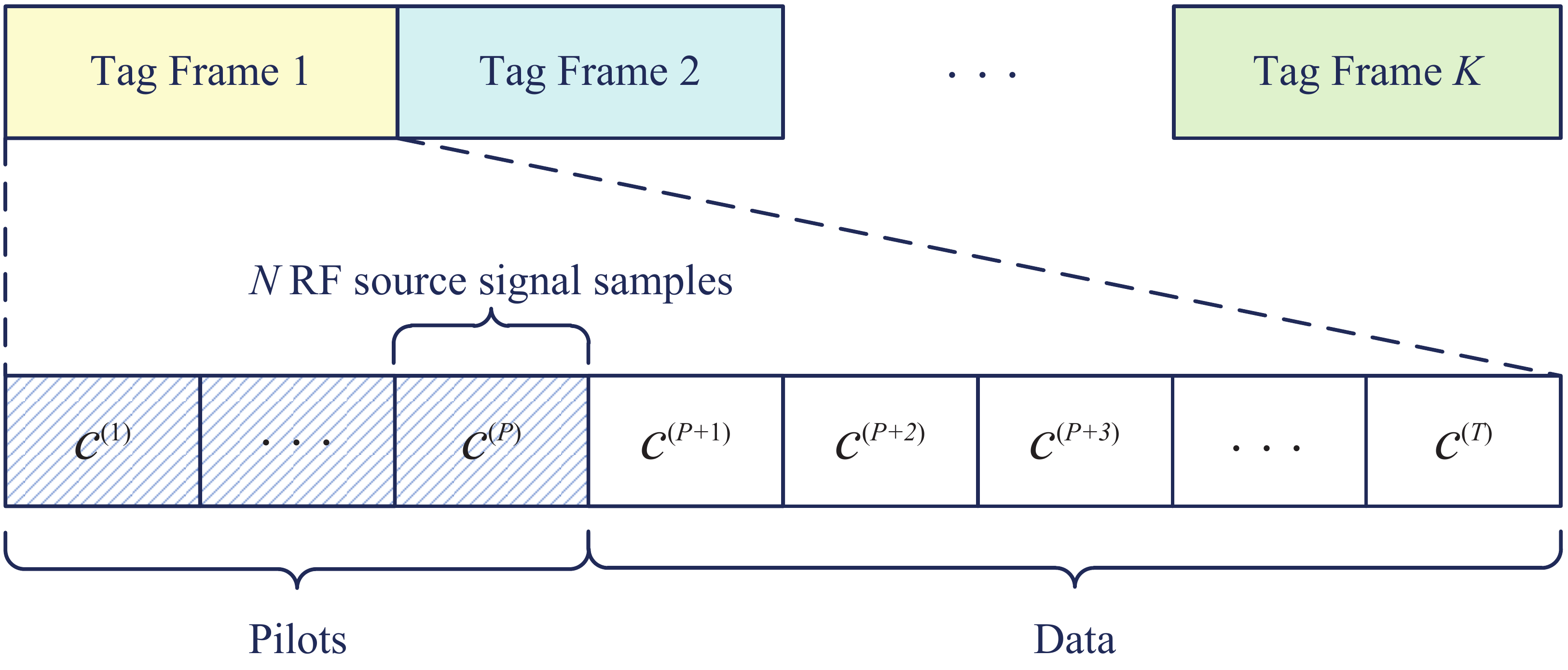}\vspace{-0.2cm}
  \caption{ The tag frame structure of the considered AmBC system. } \vspace{-0.35cm}
\end{figure}

Thus, the tag signal detection can be further formulated as a binary hypothesis testing problem:
\begin{equation}\label{sensing model}
\begin{split}
 & {{H}_{1}}:\mathbf{x}_n^{(t)} = \mathbf{w}s_n^{(t)} + \mathbf{u}_n^{(t)}, \\
 & {{H}_{0}}:\mathbf{x}_n^{(t)} = \mathbf{h}s_n^{(t)} + \mathbf{u}_n^{(t)}, \\
\end{split}
\end{equation}
where $\mathbf{w} = \mathbf{h}+\alpha f\mathbf{g}$, and $H_1$ and $H_0$ denote the hypotheses that the tag symbol $c^{(t)}=1$ and $c^{(t)}=0$, respectively.
For the ease of the following analysis, we first define the received signal-to-noise ratio (SNR) of the direct link as
\begin{equation}\label{SNR}
\mathrm{SNR} = \frac{E(||\mathbf{h}s_n^{(t)}||^2)}{E(||\mathbf{u}_n^{(t)}||^2)}.
\end{equation}
Besides, the relative coefficient between the direct signal path and the backscattered signal path is defined as a ratio of their average channel gains which is given by
\begin{equation}\label{relative_SNR}
\zeta = \frac{E(||\alpha f \mathbf{g}||^2)}{E(||\mathbf{h}||^2)}.
\end{equation}

Based on the above system model, we now introduce the optimal likelihood ratio test as a benchmark to the considered AmBC system.
According to (\ref{sensing model}), the distribution of $\mathbf{x}_n^{(t)}$ is
\begin{equation}\label{random-x(n)}
\mathbf{x}_n^{(t)}\sim
\bigg\{ \begin{matrix}
   \mathcal{C}\mathcal{N}(\mathbf{0},\mathbf{\Sigma}_1),\;{{H}_{1}} \\
   \mathcal{C}\mathcal{N}(\mathbf{0},\mathbf{\Sigma}_0),\;{{H}_{0}}   \\
\end{matrix},
\end{equation}
where $\mathbf{\Sigma}_1 = \sigma_s^2\mathbf{w}\mathbf{w}^{H}+\sigma _{u}^{2}{{\mathbf{I}}_{M}}$ and $\mathbf{\Sigma}_0 = \sigma_s^2\mathbf{h}\mathbf{h}^{H}+\sigma _{u}^{2}{{\mathbf{I}}_{M}}$. Let
$\mathbf{X}^{(t)}=[\mathbf{x}_1^{(t)},\mathbf{x}_2^{(t)},\cdots,\mathbf{x}_N^{(t)}], \forall t$,
represent a sampling matrix of the $t$-th tag symbol at the reader.
If perfect CSI, e.g., $\mathbf{w}$ and $\mathbf{h}$, are known at the reader, the logarithmic form of the optimal likelihood ratio test (LRT) under the CSCG ambient source can be derived as \cite{kay1998fundamentals}
\begin{equation}\label{L_R}
L({\mathbf{X}}^{(t)}) = \sum\limits_{n = 0}^{N - 1} \ln \left({\frac{{p\left( {\mathbf{x}_n^{(t)}|{H_1};\mathbf{0},{\mathbf{\Sigma}_1}} \right)}}{{p\left( {\mathbf{x}_n^{(t)}|{H_0};\mathbf{0},{\mathbf{\Sigma}_0}} \right)}}}\right),
\end{equation}
where
\begin{equation}\label{}
p\left( {\mathbf{x}_n^{(t)}|{H_i};\mathbf{0},{\mathbf{\Sigma} _i}} \right) = \frac{1}{{{\pi ^M}\det ({\mathbf{\Sigma} _i})}}\exp \left( { - (\mathbf{x}_n^{(t)})^H\mathbf{\Sigma} _i^{ - 1}{{\mathbf{x}_n^{(t)}}}} \right).
\end{equation}

Although the LRT can achieve the optimal BER performance, it requires the availability of perfect CSI which is not always available in practical AmBC systems due to the lack of pilot signals from the ambient RF source \cite{hu2019machine}.

\section{Covariance Matrix-based Deep Transfer Learning for Tag Signal Detection}
In this section, we propose a covariance matrix-based DTL approach to intelligently explore the features of sample covariance matrix to further improve the BER performance of AmBC systems. In the following, we will introduce the proposed covariance matrix-based neural network (CMNet) structure, the CMNet-based DTL for tag signal detection, and the CMNet-based detection algorithm, respectively.

\subsection{CMNet Structure}
As shown in Fig. 3, the CMNet consists of an input layer ($S_0$), two convolutional layers ($C_1$ and $C_2$), one pooling layer ($S_1$), one flattening layer ($C_3$), two dropout layers ($D_1$ and $D_2$), and two fully connected layers ($F_1$ and $F_2$). The convolutional, pooling, and flattening layers are used for extracting features from the input. Then, the dropout layers are added to overcome the overfitting issue which is caused by the limitation of insufficient training examples from pilots \cite{goodfellow2016deep}. Finally, the fully connected layers learn the non-linear combinations of these extracted features to further improve the performance of the task.
The corresponding hyperparameters are introduced in Table \Rmnum{1}. Here, ``ReLU'' and ``Softmax'' denote the activation functions using rectified linear unit and normalized exponential function \cite{goodfellow2016deep}, respectively. Flattening means that we flatten the last layer to create a single long feature vector. The dropout rate $\rho$ refers to the probability of training a given node in a layer.

\begin{figure}[t]
  \centering
  \includegraphics[width=0.96\linewidth]{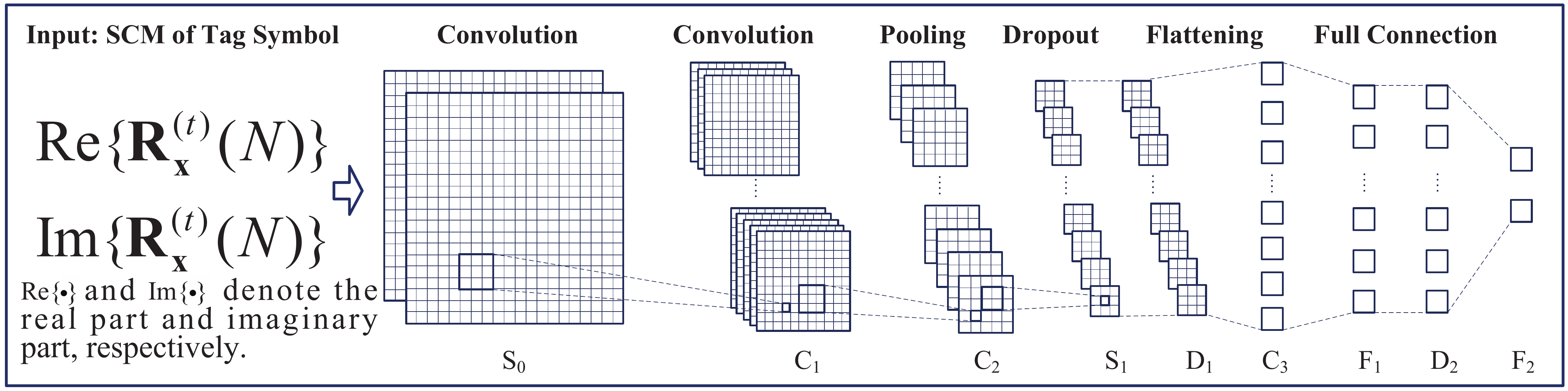}\vspace{-0.2cm}
  \caption{ The designed CMNet structure for tag signal detection. }\vspace{-0.2cm}
\end{figure}

\begin{table}[t]
\small
\caption{Hyperparameters of the proposed CMNet}
\centering
\footnotesize
\begin{tabular}{c c}
  \hline
  \vspace{-0.3cm} \\
   \multicolumn{2}{l}{\textbf{Input}: } \\
   \multicolumn{2}{l}{Sample Covariance Matrix (Dimension: $M \times M$)} \\
  \hline
  \vspace{-0.3cm} \\
  \textbf{Layer} &  \hspace{0.6cm} \textbf{Filter Size}   \\
  \hline
  \vspace{-0.3cm} \\
  ${S_0}$ & \hspace{0.6cm}  --   \\
  \hline
  \vspace{-0.3cm} \\
  ${C_1}$ + ReLU & \hspace{0.6cm} $ 64 \times ( 3 \times 3 ) $   \\
  \hline
  \vspace{-0.3cm} \\
  ${C_2}$ + ReLU & \hspace{0.6cm}  $ 64 \times ( 3 \times 3 ) $  \\
  \hline
  \vspace{-0.3cm} \\
  $S_1$ (Max-Pooling) & \hspace{0.6cm} $ 2 \times 2 $   \\
  \hline
  \vspace{-0.3cm} \\
  $C_3$ (Flattening) & \hspace{0.6cm} $ 64 \times (12 \times 12) $   \\
  \hline
  \vspace{-0.3cm} \\
  $D_1$ ($\rho = 0.5$) & \hspace{0.6cm} --  \\
  \hline
  \vspace{-0.3cm} \\
  ${F_1}$ + ReLU & \hspace{0.6cm} $ 128 \times 4608 $  \\
  \hline
  \vspace{-0.3cm} \\
  $D_2$ ($\rho = 0.25$) & \hspace{0.6cm} --  \\
  \hline
  \vspace{-0.3cm} \\
  ${F_2}$ + Softmax & \hspace{0.6cm} $ 2 \times 128 $  \\
  \hline
  \vspace{-0.3cm} \\
  \multicolumn{2}{l}{\textbf{Output}:} \\
  \multicolumn{2}{l}{Feature Vector (Dimension: $2 \times 1$)}\\
  \vspace{-0.3cm} \\
  \hline
\end{tabular}
\vspace{-0.3cm}
\end{table}

According to Fig. 2, there are $N$ RF source signal sampling periods within one tag, the sample covariance matrix (SCM) of the $t$-th tag symbol can be expressed as
\begin{equation}\label{Rx(N)}
{{\bf{R}}_{\bf{x}}^{(t)}}(N) = \frac{1}{N}\sum\limits_{n = 0}^{N - 1} {{\bf{x}}_n^{(t)}{({\bf{x}}_n^{(t)})^H}}.
\end{equation}
Considering ${{\bf{R}}_{\bf{x}}^{(t)}}(N) \in \mathbb{C}^ {M\times M}$ is a complex-valued matrix, we then adopt two different input channels to handle the the real part and imaginary part of ${{\bf{R}}_{\bf{x}}^{(t)}}(N)$, respectively.
Therefore, the CMNet can be expressed as
\begin{equation}\label{expression_CMNet}
{h}_{\theta}( {{\bf{R}}_{\bf{x}}^{(t)}}(N) ) = \left[ {\begin{array}{*{20}{c}}
{{{h}_{\theta |{H_1}}}( {{\bf{R}}_{\bf{x}}^{(t)}}(N) )}\\
{{{h}_{\theta |{H_0}}}( {{\bf{R}}_{\bf{x}}^{(t)}}(N) )}
\end{array}} \right],
\end{equation}
where ${{h}_{\theta}( \cdot )}$ is the expression of the CMNet under parameter $\theta$ and ${{h}_{\theta | {H_i}}( \cdot )}$ is the class score of $H_i$ by CMNet.

\begin{figure*}[t]
  \centering
  \includegraphics[width=5.2in, height=3in]{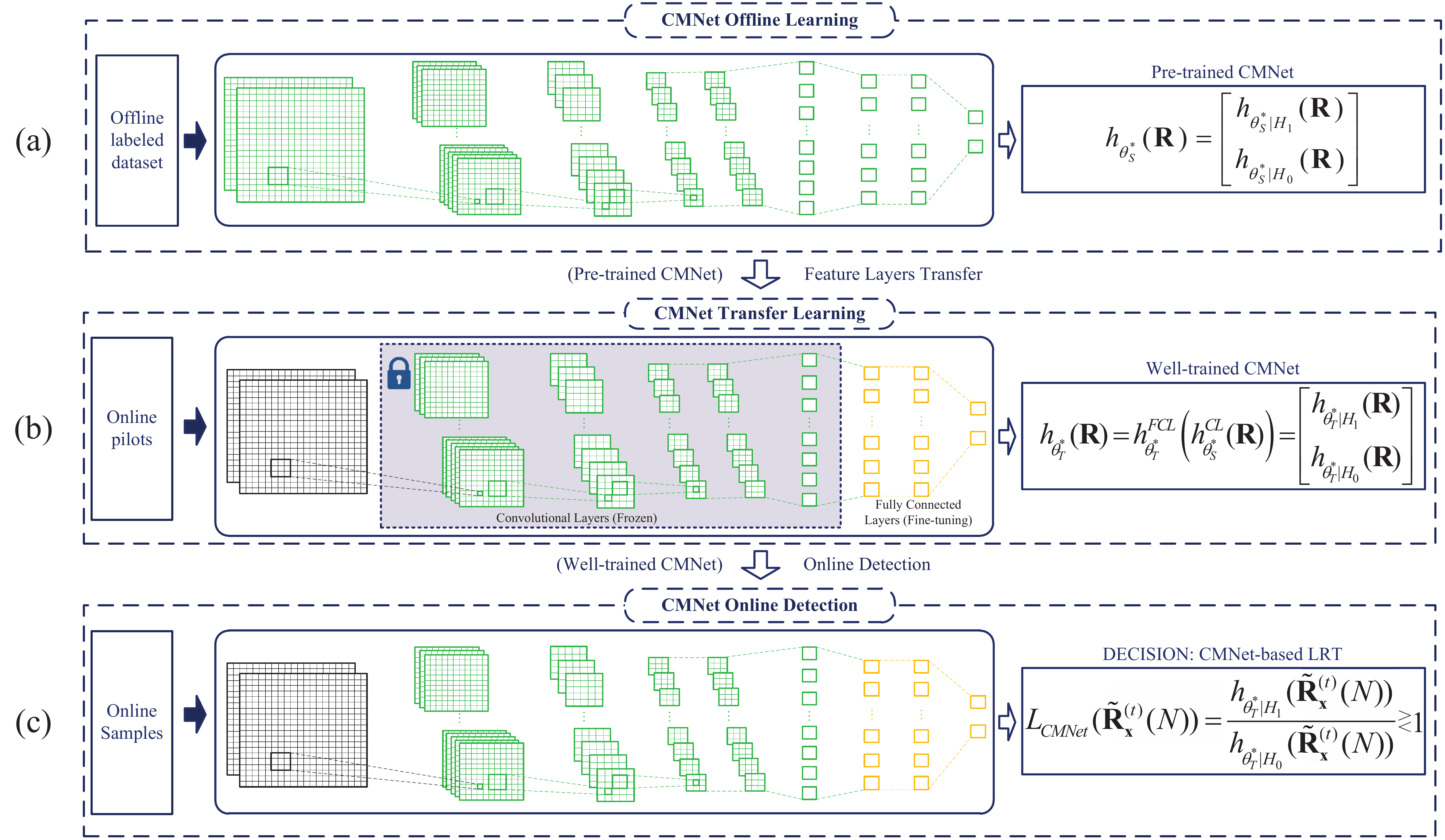}\vspace{-0.2cm}
  \caption{ The proposed CMNet-based DTL scheme for tag signal detection. }\vspace{-0.6cm}
\end{figure*}

\subsection{ CMNet-based DTL for Tag Signal Detection}
In this section, we adopt the designed CMNet to facilitate DTL for tag signal detection and propose a CMNet-based DTL scheme which consists of offline learning, transfer learning, and online detection, as shown in Fig. 4. According to \cite{pan2009survey, tan2018survey}, the training sets of offline learning and transfer learning are denoted by a source domain dataset $D_S$ and a target domain dataset $D_T$, respectively. For simplification, we assume that the channel coefficients of $D_S$ and $D_T$ are i.i.d. with distinct values. In this case, the objective of DTL for tag signal detection is to improve the BER performance of detection task in $D_T$ based on the knowledge gained from the related but different detection task in $D_S$.

Based on the above discussion, in the proposed scheme, we first establish a pre-trained CMNet to extract the common features of statistical channel models through offline learning. We then freeze the partial layers of the pre-trained CMNet and only fine-tune the remaining layers to adjust the network to the current channel coefficients through transfer learning with a few pilots. Finally, we can apply the well-trained CMNet for online detection. In the following, we will introduce the modules of offline learning (Fig. 4(a)), transfer learning (Fig. 4(b)), and online detection (Fig. 4(c)), respectively.

\subsubsection{Offline Learning}
Given $K_S$ labelled tag symbols, we adopt $\mathbf{X}_S^{(k)}=[{\mathbf{x}^{(k)}_{S_1}},{\mathbf{x}^{(k)}_{S_2}},\cdots,{\mathbf{x}^{(k)}_{S_N}}] \in \mathbb{C}^{M\times N}$ to denote the sampling matrix of the $k$-th, $k\in\{1,2,\cdots,K_S\}$, tag symbol.
Based on this, we build the source domain dataset as:
\begin{equation}\label{}
\begin{split}
D_S  =({\Omega }_S,Z_S) = &  \Big\{({{\bf{R}}_{{\bf{x}}_S}^{(1)}}(N),{{{z}}_S^{(1)}}),({{\bf{R}}_{{\bf{x}}_S}^{(2)}}(N),{{{z}}_S^{(2)}}),  \\
&\cdots ,({{\bf{R}}_{{\bf{x}}_S}^{(K_S)}}(N),{{{z}}_S^{(K_S)}})\Big\}, \\
\end{split}
\end{equation}
Here, $\Omega_S$ is the set of sample covariance matrices and ${{\bf{R}}_{{\bf{x}}_S}^{(k)}}(N) = \frac{1}{N}{\mathbf{X}_S^{(k)}}{(\mathbf{X}_S^{(k)})}^{H}$ is the $k$-th sample covariance matrix. Correspondingly, $Z_S$ is the set of tag symbols and ${{z}_S^{(k)}} \in \{1,0\}$ is the $k$-th label, where ${z}_S^{(k)} = 1$ (or $0$) refers to the hypothesis $H_1$ (or $H_0$).
According to (\ref{expression_CMNet}), the class score can be rewritten as the following probability expressions \cite{goodfellow2016deep}:
\begin{equation}\label{conditional_pro_single_example}
{h_{\theta_S|{H_i}}}({{{\bf{R}}_{{\bf{x}}_S}^{(k)}(N)}}) = P( z_S^{(k)} = i|{{{\bf{R}}_{{\bf{x}}_S}^{(k)}(N)}};\theta_S ),\\
\end{equation}
where $i \in \{1,0\}$ and $P( z_S^{(k)}|{{{\bf{R}}_{{\bf{x}}_S}^{(k)}(N)}};\theta_S )$ denotes the conditional probability under $\theta_S$.

The goal of the offline learning is to maximize the likelihood
\begin{equation}\label{}
\begin{split}
L(\theta_S )& = P(Z_S|{\Omega }_S;\theta_S ) \\
& =\prod\limits_{k = 1}^{K_S} {{{({h_{\theta_S|{H_1}} }({y_S^{(k)}}))}^{{z_S^{(k)}}}}{{( {h_{\theta_S|{H_0}} }({y_S^{(k)}}))}^{1 - {z_S^{(k)}}}}}. \\
\end{split}
\end{equation}
Therefore, we need to find parameter $\theta_S^*$ to maximize the posteriori probability $P(Z_S|{\Omega }_S)$, i.e.,
\begin{equation}\label{theta_MAP}
{\theta_S ^*} = \arg {\kern 1pt} {\kern 1pt} \mathop {\max }\limits_{\theta_S}  P( Z_S|{\Omega }_S;\theta_S ),
\end{equation}
which is equivalent to minimizing the cost function
\begin{equation}\label{CMCNNv2_cost_function_Source}
\begin{split}
{J_{\mathrm{CMNet}}}(\theta_S )  = & - \frac{1}{K_S}\sum\limits_{k = 1}^{K_S} {{z_S^{(k)}}\ln ({{h}_{\theta_S|{H_1} }}({{{\bf{R}}_{{\bf{x}}_S}^{(k)}(N)}}))}\\
&+ (1 - {z_S^{(k)}})\ln({{h}_{\theta_S|{H_0}} }({{\bf{R}}_{{\bf{x}}_S}^{(k)}(N)})). \\
\end{split}
\end{equation}
Then, exploiting the backpropagation (BP) algorithm \cite{goodfellow2016deep}, we can obtain the pre-trained CMNet:
\begin{equation}\label{fv_CMNet}
{h}_{\theta_S^*}( \mathbf{R} ) = \left[ {\begin{array}{*{20}{c}}
{{{h}_{\theta_S^* |{H_1}}}( \mathbf{R} )}\\
{{{h}_{\theta_S^* |{H_0}}}( \mathbf{R} )}
\end{array}} \right],
\end{equation}
where $\mathbf{R}$ denotes the input matrix given by (\ref{Rx(N)}) and ${{h}_{\theta_S^*}( \cdot )}$ is the expression of the pre-trained CMNet.

\subsubsection{Transfer Learning}
According to the frame structure in Fig. 2, there are $P$ pilots for transfer learning. For any pilot, we use $\mathbf{X}_T^{(k)}=[{\mathbf{x}^{(k)}_{T_1}},{\mathbf{x}^{(k)}_{T_2}},\cdots,{\mathbf{x}^{(k)}_{T_N}}]
$ to denote the sampling matrix of the $k$-th, $k\in\{1,2,\cdots,K_T\}$ tag symbol\footnotemark\footnotetext{Note that we can use data augmentation to generate $K_T \geq P$ examples based on the $P$ pilot symbols}.
We can then build the target domain dataset as:
\begin{equation}\label{}
\begin{split}
D_T  =({\Omega }_T,Z_T) = &  \Big\{({{\bf{R}}_{{\bf{x}}_T}^{(1)}}(N),{{{z}}_T^{(1)}}),({{\bf{R}}_{{\bf{x}}_T}^{(2)}}(N),{{{z}}_T^{(2)}}),  \\
&\cdots ,({{\bf{R}}_{{\bf{x}}_T}^{(K_T)}}(N),{{{z}}_T^{(K_T)}})\Big\}. \\
\end{split}
\end{equation}
Here, $\Omega_T$ is the set of sample covariance matrices and ${{\bf{R}}_{{\bf{x}}_T}^{(k)}}(N) = \frac{1}{N}{\mathbf{X}_T^{(k)}}{(\mathbf{X}_T^{(k)})}^{H}$ is the $k$-th sample covariance matrix. Correspondingly, $Z_T$ is the set of tag symbols and ${{z}_T^{(k)}} \in \{1,0\}$, $\forall k$, is the label.

Similar to (\ref{CMCNNv2_cost_function_Source}), the cost function for transfer learning is
\begin{equation}\label{CMCNNv2_cost_function_Target}
\begin{split}
{J_{\mathrm{CMNet}}}(\theta_T ) =&  - \frac{1}{K_T}\sum\limits_{k = 1}^{K_T} {{z_T^{(k)}}\ln ({{h}_{\theta_T|{H_1} }}({{{\bf{R}}_{{\bf{x}}_T}^{(k)}(N)}}))}\\
& + (1 - {z_T^{(k)}})\ln({{h}_{\theta_T|{H_0}} }({{\bf{R}}_{{\bf{x}}_T}^{(k)}(N)})).\\
\end{split}
\end{equation}
As shown in Fig. 4(b), during the training process, we can reuse the convolutional layers of the pre-trained CMNet, i.e., freezing the convolutional layers and only update the parameters of the fully connected layers applying the BP algorithm. Finally, we can obtain the well-trained CMNet:
\begin{equation}\label{fv_function}
h_{\theta_T^*}( \mathbf{R} ) = h_{{\theta _T^*}}^{\mathrm{FCL}}(h_{\theta _S^*}^{\mathrm{CL}}(\mathbf{R})) = \left[ {\begin{array}{*{20}{c}}
{{h_{\theta_T^*|{H_1}}}( \mathbf{R} )}\\
{{h_{\theta_T^*|{H_0}}}( \mathbf{R} )}
\end{array}} \right],
\end{equation}
where $h_{\theta_T^*}( \cdot )$ denotes the expression of the well-trained CMNet with the well-trained parameter $\theta_T^*$, $h_{{\theta _T^*}}^{\mathrm{FCL}}(\cdot)$ denotes the fully connected layers ($F_1\rightarrow F_2$) obtained through fine-tuning, and $f_{\theta _S^*}^{\mathrm{CL}}(\cdot)$ represents the convolutional layers ($C_1\rightarrow C_3$) obtained from the pre-trained CMNet.

From a probabilistic viewpoint, we can then rewrite the outputs of the well-trained CMNet as
\begin{equation}\label{}
\begin{split}
{H_1}: h_{\theta_T^*|{H_1}}(\mathbf{R}) = P({H_1}|\mathbf{R}), \\
{H_0}: h_{\theta_T^*|{H_0}}(\mathbf{R}) = P({H_0}|\mathbf{R}), \\
\end{split}
\end{equation}
where $P({H_i}|\mathbf{R})$ denotes the posterior probability expression.
Based on Bayes' theorem, we can obtain the likelihood expressions as
\begin{equation}\label{ConPro_Hi}
L({H_i}|\mathbf{R}) = \frac{{P({H_i}|\mathbf{R})}\cdot P(\mathbf{R})}{{P({H_i})}} = \frac{{h_{\theta_T^* |{H_i}}^*(\mathbf{R})}\cdot P(\mathbf{R})}{{P({H_i})}},
\end{equation}
where $P(H_i)$ is the priori probability of $H_i$, and $P(\mathbf{R})$ is the marginal probability of the input matrix. Note that we always set $P(H_1)=P(H_0)=0.5$ for binary communication systems. Therefore, according to the minimum error probability (MEP) criterion, we can then derive the CMNet-LRT:
\begin{equation}\label{CMNet-LRT}
{L}_{\mathrm{CMNet}}(\mathbf{R}) = \frac{L({H_1}|\mathbf{R})}{L({H_0}|\mathbf{R})} = \frac{h_{\theta_T^*|{H_1}}(\mathbf{R})}{h_{\theta_T^*|{H_0}}(\mathbf{R})} \gtrless 1,
\end{equation}
where we make a decision that $H_1$ holds if ${L}_{\mathrm{CMNet}}(\mathbf{R}) > 1$, otherwise, $H_0$ holds.

\subsubsection{Online Detection}
Given the $t$-th tag symbol's sampling matrix for detection, denoted by $\mathbf{\tilde{X}}^{(t)}=[\mathbf{\tilde{x}}_1^{(t)},\mathbf{\tilde{x}}_2^{(t)},\cdots,\mathbf{\tilde{x}}_N^{(t)}]$, the corresponding sample covariance matrix is $\mathbf{\tilde{R}_x}^{(t)}(N) = \frac{1}{N}\mathbf{\tilde{X}}^{(t)}(\mathbf{\tilde{X}}^{(t)})^{H}$. The decision of the CMNet-LRT is given by:
\begin{equation}\label{L-CMNet}
{L}_{\mathrm{CMNet}}(\mathbf{\tilde{R}_x}^{(t)}(N)) = \frac{h_{\theta_T^*|{H_1}}(\mathbf{\tilde{R}_x}^{(t)}(N))}{h_{\theta_T^*|{H_0}}(\mathbf{\tilde{R}_x}^{(t)}(N))} \mathop \gtrless\limits_{{c^{(t)}=0}}^{{c^{(t)}=1}} 1,
\end{equation}
where $c^{(t)}=1$ if ${L}_{\mathrm{CMNet}}(\mathbf{\tilde{R}_x}^{(t)}(N))>1$, otherwise, $c^{(t)}=0$.

\subsubsection{Algorithm Steps}
Based on the analysis above, a novel CMNet-based detection algorithm is proposed in \textbf{Algorithm 1}, where $i_S$ and $i_T$ represent iteration indices, and $I_S$ and $I_T$ indicate the maximum numbers of iterations of the offline learning and the transfer learning, respectively.

\begin{table}[t]
\small
\centering
\begin{tabular}{l}
\toprule[1.8pt] \vspace{-0.4cm}\\
\hspace{-0.1cm} \textbf{Algorithm 1} \hspace{0.6cm} {CMNet-based Detection Algorithm}  \\
\toprule[1.8pt] \vspace{-0.3cm}\\
\textbf{Initialization:}
   $i_S = 0$, $i_T = 0$, $I_S = a$, $I_T = b$ \\
\textbf{Offline Learning:} \\
1:\hspace{0.75cm}\textbf{Input:} Training set $D_S =({\Omega }_S,Z_S)$\\
2:\hspace{1.1cm}\textbf{while} $i_S \leq I_S $ \textbf{do} \\
3:\hspace{1.6cm}update $\theta_S$ by BP algorithm on $J_{\mathrm{CMNet}}(\theta_S)$ \\
\hspace{1.8cm} $i_S = i_S + 1$  \\
4:\hspace{1.1cm}\textbf{end while} \\
5:\hspace{0.75cm}\textbf{Output}:  ${h}_{\theta_S^*}( \cdot ) $\\
\textbf{Transfer Learning:} \\
6:\hspace{0.75cm}\textbf{Input:} Training set $D_T =({\Omega }_T,Z_T)$\\
7:\hspace{1.1cm}\textbf{while} $i_T \leq I_T $ \textbf{do} \\
8:\hspace{1.6cm}update $\theta_T$ by BP algorithm on $J_{\mathrm{CMNet}}(\theta_T)$ \\
\hspace{1.8cm} $i_T = i_T + 1$  \\
9:\hspace{1.1cm}\textbf{end while} \\
10:\hspace{0.6cm}\textbf{Output:}  ${h}_{\theta_T^*}( \cdot ) $\\
\textbf{Online Detection:} \\
11:\hspace{0.6cm}\textbf{Input:} Real-time Test data $\mathbf{\tilde{X}}^{(t)}$ \\
12:\hspace{0.95cm}\textbf{do} CMNet-LRT in (\ref{L-CMNet}) \\
13:\hspace{0.6cm}\textbf{Output:} Decision: $c^{(t)}=1$ or $c^{(t)}=0$ \\
\bottomrule[1.8pt]
\end{tabular}\vspace{-0.5cm}
\end{table}

\section{Simulation Results}
This section presents simulation results to verify the efficiency of the proposed detection method under a classical AmBC system as shown in Fig. 1. In the simulation, a CSCG ambient source is adopted, the number of antennas at the reader is $M = 16$, the number of pilots is set as $P = 10$. The STR and the length of framework are set as $N = 50$ and $NT = 5,000$, respectively. All the channel coefficients follow Rayleigh channel model and remain unchanged within each tag frame.
To evaluate the BER performance, we compare the proposed CMNet method with other three related algorithms, i.e., the optimal LRT method \cite{kay1998fundamentals}, the ED method with perfect CSI\cite{qian2017semi}, and the SVM-based method \cite{hu2019machine}. The hyperparameters of the CMNet method are shown in Table \Rmnum{1} and we set $I_S = 30$, $I_T = 60$ for Algorithm 1. For the training datasets, we set $K_S = 60,000$ and $K_T = 2,000$. All the simulation results are obtained by averaging over $100,000$ Monte Carlo realizations.

We first present the BER curves with different SNRs in Fig. 5. It is shown that although the SVM-based method achieves similar BER performance than that of the ED method with perfect CSI, there is still a large gap compared to that of the optimal LRT method. In contrast to them, the BER performance of the proposed CMNet method approaches closely to the optimal performance achieved by the LRT method with perfect CSI.
It is worth mentioning that the proposed CMNet method achieves a SNR gain of 4 dB at BER $\,\approx10^{-2}$ compared to the traditional SVM-based method. The reason is that the SVM-based method makes decisions depending on the sole energy-based feature of the received signals, while the proposed CMNet method makes decisions by exploiting the discriminative features from the sample covariance matrices in a data-driven approach.

Fig. 6 presents the curves of BER versus relative coefficients. It is shown that when the value of relative coefficient $\zeta$ increases, the BER of each method decreases gradually. This is because the increase of $\zeta$ improves the strength of the reflected path, which makes the tag signals easier to be distinguished against the signals of the direct path. In addition, we can find that the proposed CMNet method outperforms both the SVM and ED methods dramatically, achieving almost the same optimal performance as the LRT detector.

\begin{figure}[t]
  \centering
  \includegraphics[width=2.5in,height=2in]{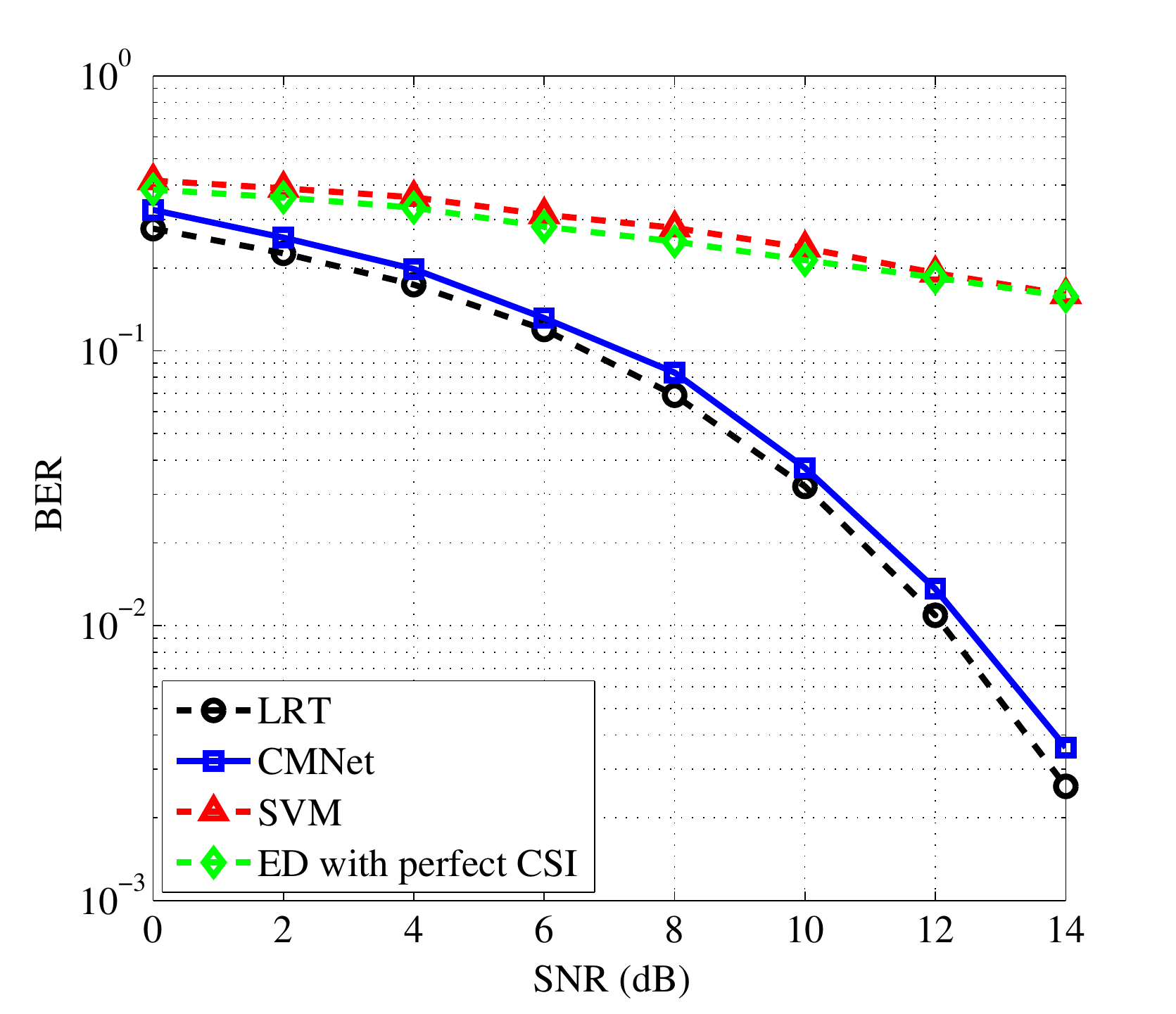}\vspace{-0.4cm}
  \caption{BER curves with different SNRs under $\zeta$ = -30 dB.}\vspace{-0.4cm}
\end{figure}

\begin{figure}[t]
  \centering
  \includegraphics[width=2.5in,height=2in]{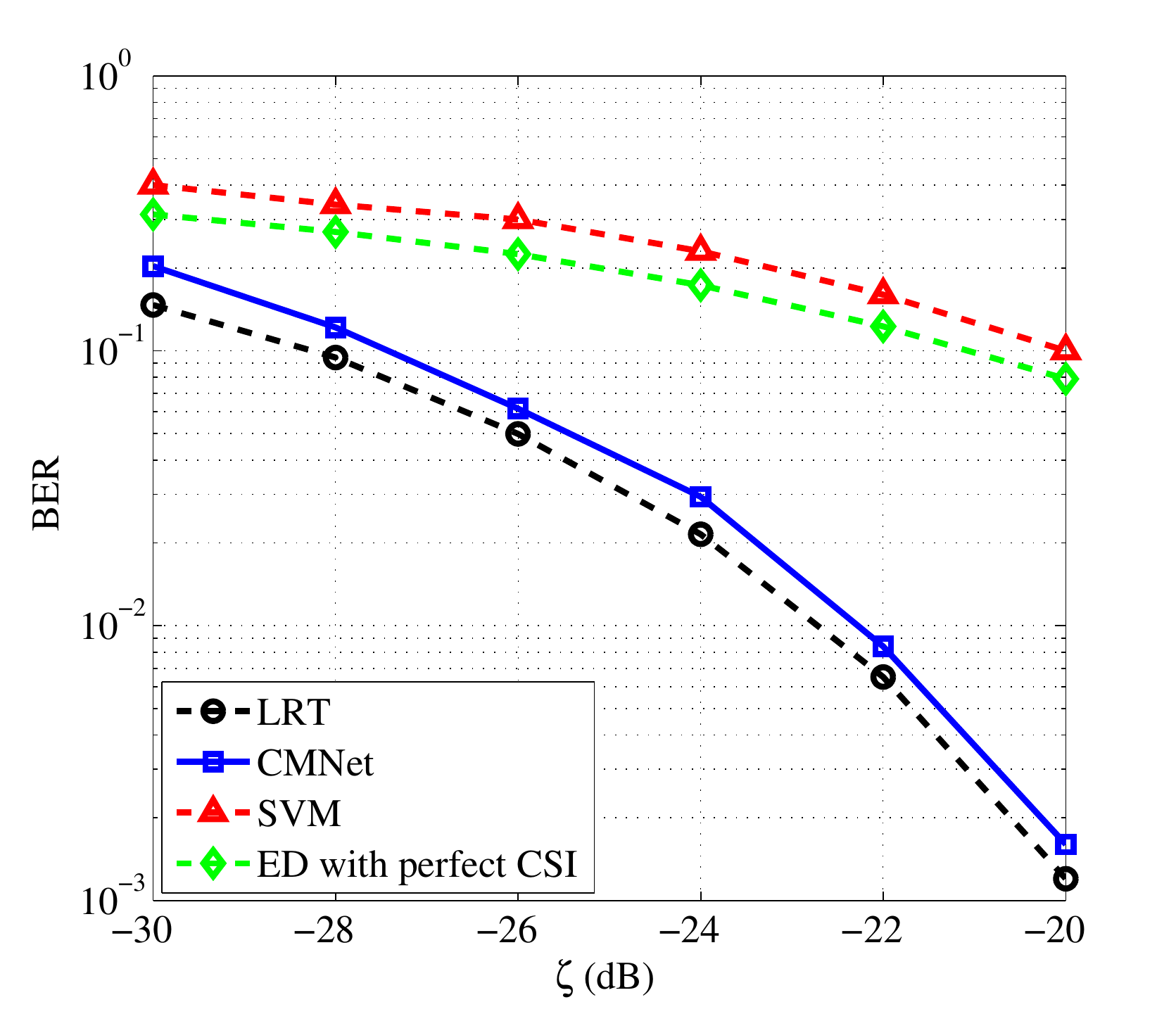}\vspace{-0.4cm}
  \caption{BER curves with different relative coefficients under SNR = 5 dB.}\vspace{-0.4cm}
\end{figure}

Fig. 7 shows the BER curves with different numbers of antennas.
It is shown that both the SVM and ED methods improve the performance slowly, while the performance of the CMNet method scales with the numbers of antennas with the same slope as the optimal LRT method. This is because the proposed method can efficiently exploit the spatial degrees of freedom offered by the antennas and it can always exploit distinguishable features for improving the system performance.

\begin{figure}[t]
  \centering
  \includegraphics[width=2.5in,height=2in]{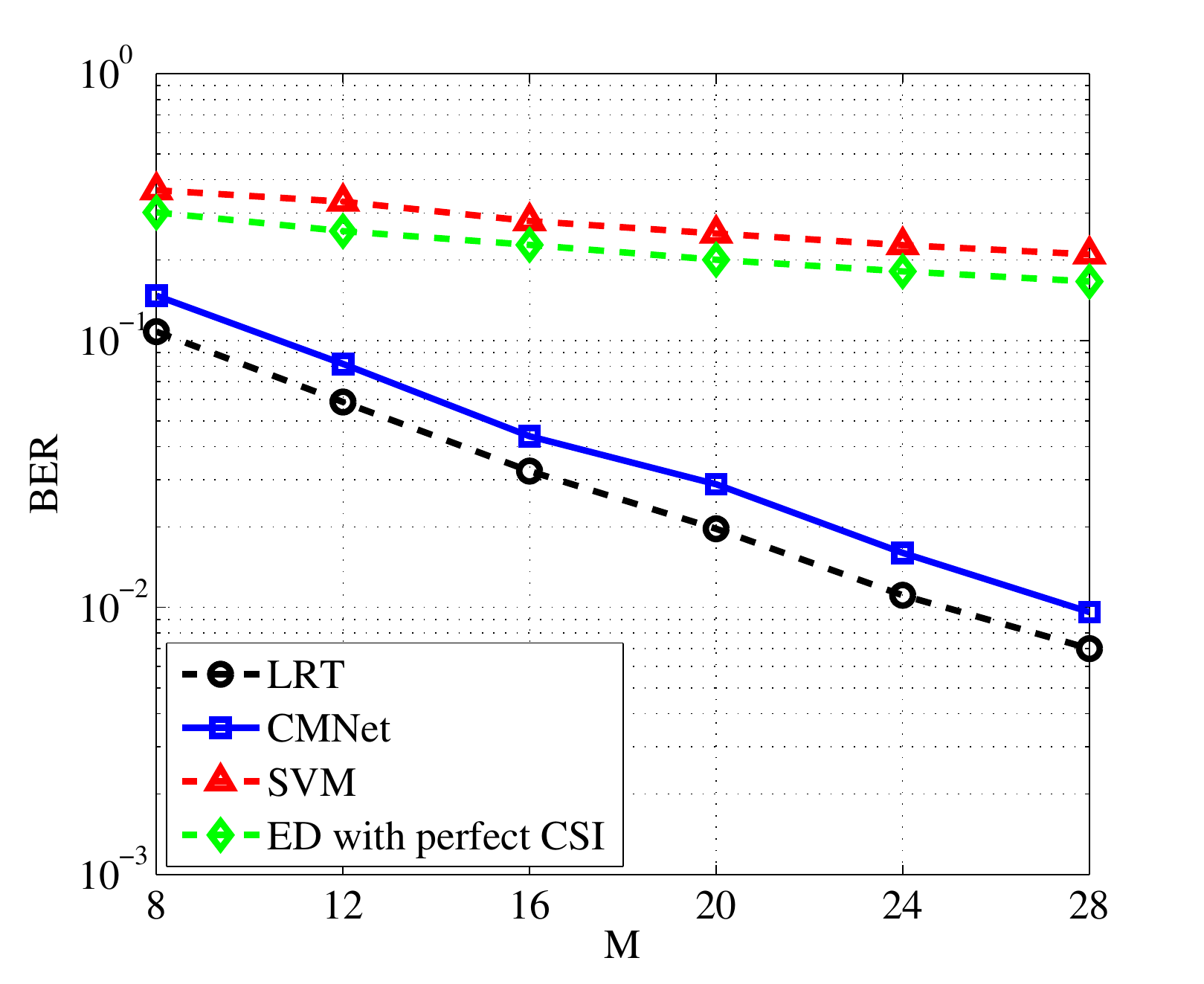}\vspace{-0.4cm}
  \caption{BER curves with different numbers of antennas under SNR = 10 dB and $\zeta$ = -30 dB.}\vspace{-0.4cm}
\end{figure}

\section{Conclusions}
This paper developed a DTL-based tag signal detection approach to implicitly extract the features of channel and directly recover tag symbols. To efficiently capture the time-varying features of wireless environment, a novel CMNet was designed to exploit the discriminative features of sample covariance matrix through offline pre-training and online fine-tuning with a few pilots in real-time. Specifically, a CMNet-LRT was derived for tag signal detection, which enables the design of an effective detector.
Extensive simulation results showed that the proposed CMNet method can achieve a close-to-optimal performance without explicitly obtaining the CSI.

\bibliographystyle{ieeetr}

\setlength{\baselineskip}{10pt}

\bibliography{ReferenceSCI2}

\end{document}